\documentclass[aps,amssymb,floatfix,prd,amsmath,preprintnumbers]{revtex4}
\usepackage{epstopdf}
\usepackage{capt-of}
\usepackage{xcolor}
\usepackage{graphicx}  
\usepackage{dcolumn}   
\usepackage{bm}
\begin{document}
\input epsf.tex

\title{\bf Magnetized cosmological model with variable deceleration parameter}

\author{Sankarsan Tarai, \footnote{Department of Mathematics, National Institute of Technology Calicut, Kozhikode, Kerala, India-673601, E-mail:tsankarsan87@gmail.com} Fakhereh Md. Esmaeili \footnote{School of Physics, University of Hyderabad , Hyderabad-500046, India, E-mail:astrosat92@gmail.com}, B. Mishra, \footnote{Department of Mathematics, Birla Institute of Technology and Science-Pilani, Hyderabad Campus, Hyderabad-500078, India, E-mail:bivudutta@yahoo.com}, S.K. Tripathy \footnote{Department of Physics, Indira Gandhi Institute of Technology, Sarang, Dhenkanal, Odisha-759146, India, E-mail:tripathy\_sunil@rediffmail.com}
}

\affiliation{ }

\begin{abstract}
\begin{center}
\textbf{Abstract}
\end{center}
In this paper, we have derived the field equations in an extended theory of gravity in an anisotropic space time background and in the  presence of magnetic field. The physical and geometrical parameters of the models are determined with respect to the  Hubble parameter using some algebraic approaches. A time varying scale factor has been introduced to analyze the behavior of the model. From some diagnostic approach, we found that the model behaves as  $\Lambda_{CDM}$ model at late time of cosmic evolution. 
\end{abstract}

\maketitle
\textbf{PACS number}: 04.50kd.\\
\textbf{Keywords}:  Extended gravity; Magnetic field; Energy conditions; Geometrical diagnostic.

\section{Introduction} 

Theoretically, researchers found two convenient answers to the observational claim on the accelerated expansion of the universe. The first one is the most common dark energy  approach. The assumption here is that the universe filled with mysterious and unknown form of energy which is  responsible for its mass energy balance as well causing the accelerated expansion \cite{Caldwell98,Padmanabhan03,Peebles03,Tsujikawa13}. The other approach is the dark gravity approach, where it is assumed that at large scales, the gravitational force may have a very different behavior as compared to standard general theory of relativity. Further there are two ways to this dark gravity approach, (i) modification of the geometric part of the Einstein-Hilbert Lagrangian only; and (ii) maximal extensions of Einstein-Hilbert action such that both geometric and matter term in Einstein-Hilbert action can be modified. The common example of approach (i) is the $f(R)$ gravity  \cite{Carroll04,Nojiri07,Sotiriou10} and for (ii) the $f(R,T)$ gravity  \cite{Bertolami07,Harko11}.\\

After this formulation, plethora of cosmological models with $f(R,T)$  gravity are presented on the accelerated expansion in different matter form such as perfect fluid, viscous fluid, cosmic string, electromagnetic field etc. Shabani and Ziaie \cite{Shabani18}, Tripathy et al. \cite{Tripathy19}, Mishra et al. \cite{Mishra19},  Tripathy and Mishra \cite{Tripathy20} have studied the bouncing cosmological models. Dynamical behaviors of the models with the known scale factors such as power law, exponential law, hybrid scale factor and hyperbolic scale factor are also presented to study the expanding behavior of the universe \cite{Mishra18a,Arbuzova18,Mishra18b,Esmaeili18,Mishra18c}. Stability of the models under this gravity is another aspect to validate the modified gravity \cite{Shabani17,Sharif19,Zubair19,Bhatti20}. Bulk viscous model has been studied extensively in $f(R,T)$ gravity. Mahanta \cite{Mahanta14} presented the effective stiff fluid and viscous fluid in LRS Bianchi I space-time. Singh and Kumar \cite{Singh14} have compared the viscous model with the non-viscous one with the possible occurrence of big rip.  Mishra et al. \cite{Mishra18d} presented the dynamical aspects of the model with the viscous fluid with power law expansion. Debnath \cite{Debnath19} studied the model in presence of bulk viscosity which are permitted in theories like Full Israel Stewart theory, Truncated Israel Stewart theory and Eckart theory.  Wu et al. \cite{Wu18} have formulated Palatini formulation of $f(R,T)$ gravity and have presented its cosmological implications. Godani and Samanta \cite{Godani19} have studied the wormhole solutions and have shown that the presence of exotic matter is minimized. Moraes et al. \cite{Moraes17} have obtained the energy conditions of static wormholes and found the physical and geometrical solutions in an analytic approach. Deb et al. \cite{Deb19} have presented a specific model of anisotropic strange stars in the modified $f(R,T)$ gravity whereas Biswas et al. \cite{Biswas20}  have studied the anisotropic spherically symmetric strange star under the background of $f(R,T)$ gravity. \\

Presence of magnetic field in the matter term have an important role at early time of the cosmic dynamics description \cite{Maden89,King07}. Jacobs \cite{Jacobs69} has claimed that at early time of evolution, the magnetic field produced large expansion anisotropies during the radiation dominated era and the origin and evolution of galactic magnetic fields are still not known.  The perturbation due to primordial magnetic field induces temperature and polarization anisotropies in cosmic microwave background radiation (CMB) and therefore large scale magnetic fields can be detected by observing their effects on the CMB radiation \cite{Chen04,Kahniashvili08,Bernui08}. A magnetic field with an amplitude of $10^{-8}$-$10^{-9}$ Gauss may leave traces on CMB radiation. It is assumed that the observed galactic magnetic fields were originated from cosmological or astrophysical seeds and possibly generated during the recombination era, i.e $T>0.25$ eV through the mechanism of magnetogenesis during inflationary epoch \cite{Turner1988, Ratra1992}.  Cosmological phase transitions could also produce primordial magnetic fields \cite{Vachaspati1991, Cornwall1997}. The origin of cosmic magnetic fields can be attributed to primordial quantum fluctuations and their seeds may be in the range $10^{-18}$-$10^{-27}$ Gauss or less \cite{Grasso01,Giovannini04,Giovannini08,Andrianov08}. One may refer to the reviews on the origin of primordial magnetic field \cite{Widrow2002, Barrow2007, Subramanian2016, Subramanian2019} and references therein.\\

Several aspects of electromagnetic field have been studied in $f(R,T)$ gravity. In this modified theory of gravity, Bamba et al. \cite{Bamba12} have studied the logarithmic non minimal gravitational coupling of the electromagnetic theory. Also, coupled with electromagnetic field, Mazharimousavi et al. \cite{Mazhar12} have obtained solution for modified theory of gravity. Sharma and Singh \cite{Sharma14} studied the non-diagonal anisotropic metric based cosmology in magnetic field. Ram and Chandel \cite{Ram15} have derived and analysed the cosmological solutions of massive string in presence of magnetic field. Yousaf and Bhatti \cite{Yousaf16} have investigated the role of electromagnetic field and $f(R)$ corrections on the evolution of cylindrical compact object. The magnetized strange quark matter solution in FRW metric has been obtained by Aktas and Aygun \cite{Aktas17}. Pradhan and Jaiswal have \cite{Pradhan18} compared the massive string cosmological model with the recent observationally obtained data.  Islam and Basu \cite{Islam18} have studied the interior solutions of distributions of magnetized fluid inside a sphere. Tarai and Mishra \cite{Tarai18} have shown the presence of magnetic field in the matter field substantially effect the dynamical behavior of the model. Khan and Khan \cite{Khan19} have pointed that electromagnetic field diminishes the bound of $f(R,T)$ term by diminishing the pressure and hence claimed that the whole collapsing process accelerates.\\
     
Since Einstein's GR encountered with certain difficulties in describing                                                                                                              late time cosmic phenomena issue, with a changed geometry the gravitational action for the modified theory of gravity is considered as \cite{Harko11},
\begin{equation}\label{eq:1}
S=\frac{1}{16\pi}\int \sqrt{-g}f(R,T)d^{4}x + \int \sqrt{-g}\mathcal{L}_{m} d^{4}x.
\end{equation}

Here, $f(R,T)$ is an arbitrary function of the Ricci scalar $R$ and $T$ is the trace of energy momentum tensor. The matter Lagrangian $\mathcal{L}_m$ describes the matter contribution. In the literature, two approaches are suggested to derive the field equations of $f(R,T)$ gravity (i) with the standard metric formalism, by varying the action with respect to the metric tensor, where the affine connection depends on metric tensor and (ii) the Palatini formalism, where the metric tensor and affine connection treated independent while varying the action. We consider here $\mathcal{L}_m=-p$ and used the first approach, where affine connection depends on metric tensor. Among three cases suggested for the functional $f(R,T)$ (Ref.\cite{Harko11}), we have considered the non-minimal matter-geometry coupling case, $f(R,T)=f_1(R)+f_2(T)$. By varying the action \eqref{eq:1} with respect to the metric tensor $g_{\mu \nu}$, the field equations of $f(R,T)$ gravity can be expressed as,

\begin{equation}\label{eq:2}
f_R (R) R_{\mu\nu}-\frac{1}{2}f(R)g_{\mu\nu}-\left(\nabla_{\mu} \nabla_{\nu}-g_{\mu\nu}\Box\right)f_R(R)=8\pi T_{\mu\nu}+f_T(T)T_{\mu\nu}+\left[f_T(T)p+\frac{1}{2}f(T)\right]g_{\mu\nu}
\end{equation}

where, $g_{\mu \nu}$ is the gravitational metric potential. $f_R=\frac{\partial f_1(R)}{\partial R}$ and $f_T=\frac{\partial f_2(T)}{\partial T}$ are the respective partial differentiations. $T$ is the trace of the energy momentum tensor $T_{\mu\nu}$, which has been described here as the combination of perfect fluid and electromagnetic field in the form 

\begin{equation} \label{eq:3}
T_{\mu\nu}=(\rho+p)u_{\mu}u_{\nu}-pg_{\mu\nu}+ E_{\mu \nu}
\end{equation}  

where, $\rho$ and $p$ respectively denote the matter energy density and pressure. $u^{\mu}u_{\mu}=1$ and $u^{\mu}$ is the four velocity vector fluid in a co-moving coordinate system. $E_{\mu \nu}$ represents the electromagnetic field and can be defined as, 
\begin{equation}  \label{eq:4}
E_{\mu \nu}=\frac{1}{4 \pi} \left[ g^{sp}F_{\mu s}F_{\nu p}-\frac{1}{4} g_{\mu \nu}F_{sp}F^{sp}\right].
\end{equation}

In eqn. \eqref{eq:4}, $F_{sp}$ represents the electromagnetic field tensor. We are interested to study the effect of magnetic field on the given model and therefore, we wish to assume an infinite electrical conductivity which ensures only the non-vanishing magnetic components of the antisymmetric electromagnetic field tensor $F_{\mu \nu}$.  In otherwords, we have $F_{14}=F_{24}=F_{34}=0$. Also, the magnetic field is assumed to align along the $x$-axis, so that the only non-vanishing component of electromagnetic field tensor is $F_{23}$. From Maxwell's equations, we get $F_{23}=-F_{32}=\mathcal{M}$.

We wish to consider a functional $f(R,T)$ for which the field equations can be reduced to the usual GR field equations. A popular choice is $f(R,T)=R+2f(T)$ \cite{Das2017, Shamir2015,Moraes2016b, Tripathy19,Mishra18a,SKT2020}. Following a recent work \cite{Mishra18c}, we have considered here $f_1(R)=R$ and $f_2(T)=\beta T+2\Lambda_0$, where $\beta$ be the coupling constant and $\Lambda_0$ be the standard cosmological constant. The motivation behind such a choice is that, in the event of a vanishing coupling constant $\beta$, the gravity theory reduces to that of GR with a cosmological constant. The field eqns. \eqref{eq:2} can now be expressed as

\begin{equation}\label{eq:5}
G_{\mu \nu}= (8\pi+2\beta)T_{\mu\nu}+\Lambda(T)g_{\mu\nu},
\end{equation}
where, $\Lambda(T)= (2p+T)\beta+\Lambda_{0}$ as the time variable cosmological constant. \\

In Sec. II, the field equations of the extended theory of gravity derived in an anisotropic space-time and the physical parameters are derived in the form of Hubble parameter. In Sec. III, the physical parameters are derived with scale factor of variable deceleration parameter. More geometrical analysis and energy conditions are done in Sec. IV and the conclusion and key points are presented in Sec. V. Here, we have considered the natural system of units with $G = c = 1$, where $G$ is the Newtonian gravitational constant and $c$ is the speed of light in vacuum.
\section{Field Equations and Physical Parameters}
We have considered Bianchi $VI_h$ space-time in the form
\begin{equation}\label{eq:6}
ds^{2}=dt^{2}-\phi^{2}dx^{2}-\varphi^{2}e^{2x}dy^{2}-\psi^{2}e^{2hx}dz^{2},
\end{equation}
where $\phi$, $\varphi$, $\psi$ are function of the cosmic time $t$.  The exponent $h$ above is self-governing constant and can take integral values in the range $ h=[-1,1]$. In this work, we have considered the self-governing constant $h=-1$. Tripathy et al. [21] have studied the importance of $h=-1$ on the dynamics of the universe and its significance in the study of general relativity and relativistic astrophysics. Now, the field eqns. \eqref{eq:5} give the following set of equations,
\begin{eqnarray}\label{eq:7}
\frac{\ddot{\varphi}}{\varphi}+\frac{\ddot{\psi}}{\psi}+\frac{\dot{\varphi}\dot{\psi}}{\varphi\psi}+\frac{1}{\phi^{2}}&=&-\alpha(p+\mathcal{H})+\beta(\rho-p)+\Lambda_{0}\\
\frac{\ddot{\phi}}{\phi}+\frac{\ddot{\psi}}{\psi}+\frac{\dot{\phi}\dot{\psi}}{\phi\psi}-\frac{1}{\phi^{2}}&=&-\alpha(p-\mathcal{H})+\beta(\rho-p)+\Lambda_{0} \\ \label{eq:8}
\frac{\ddot{\phi}}{\phi}+\frac{\ddot{\varphi}}{\varphi}+\frac{\dot{\phi}\dot{\varphi}}{\phi\varphi}-\frac{1}{\phi^{2}}&=&-\alpha(p-\mathcal{H})+\beta(\rho-p)+\Lambda_{0} \\\label{eq:9}
\frac{\dot{\phi}\dot{\varphi}}{\phi\varphi}+\frac{\dot{\varphi}\dot{\psi}}{\varphi\psi}+\frac{\dot{\phi}\dot{\psi}}{\phi\psi}-\frac{1}{\phi^{2}}&=&\alpha(\rho-\mathcal{H})+\beta(\rho-p)+\Lambda_{0} \\\label{eq:10}
\frac{\dot{\varphi}}{\varphi}-\frac{\dot{\psi}}{\psi}&=& 0.\label{eq:11}
\end{eqnarray}
Eqn.\eqref{eq:11} gives $\varphi=\psi$, by suitably absorbing the integrating constant. An overdot in the field variable denotes the ordinary time derivative. $\alpha=8\pi+2\beta$ be the revised coupling constant and $\mathcal{H}=\frac{\mathcal{M}^{2}}{8\pi \varphi^{2}\psi^{2}}$ is the magnetic energy density.\\

We are interested to express the physical parameters with respect to the Hubble parameter $H$, so that the background cosmology can be studied. Now, the mean Hubble parameter, $H=\frac{\dot{\mathcal{R}}}{R}=\frac{H_x+H_y+H_z}{3}$, where $\mathcal{R}$ is the scale factor and $H_x=\frac{\dot{\phi}}{\phi}$, $H_y=\frac{\dot{\varphi}}{\varphi}$, $H_z=\frac{\dot{\psi}}{\psi}$ are the directional Hubble rates along the rectangular coordinate axis respectively. Subsequently eqn. \eqref{eq:11} gives, $H_y=H_z$. We have incorporated some amount of anistropy among the field variable in the form $H_x=kH_z$, where $k\neq1$. Now incorporating all these form into the field eqns. \eqref{eq:7}-\eqref{eq:11}, we can obtain the scale factor transformed field equations as,

\begin{eqnarray}
6(k+2)\frac{\ddot{\mathcal{R}}}{\mathcal{R}}+3(5-2k)\frac{\dot{\mathcal{R}}^{2}}{\mathcal{R}^{2}}+(k+2)^{2} \phi^{\frac{-6k}{k+2}}&=& (k+2)^2\times\left[-\alpha(p+\mathcal{H})+\beta(\rho-p)+\Lambda_{0}\right]  \label{eq:12} \\
3(k^{2}+3k+2)\frac{\ddot{\mathcal{R}}}{\mathcal{R}}+3(2k^{2}+1)\frac{\dot{\mathcal{R}}^{2}}{\mathcal{R}^{2}}-(k+2)^{2} \phi^{\frac{-6k}{k+2}} &=& (k+2)^2\times \left[-\alpha(p-\mathcal{H})+\beta(\rho-p)+\Lambda_{0} \right] \label{eq:13} \\ 
9(2k+1)\frac{\dot{\mathcal{R}}^{2}}{\mathcal{R}^{2}}-(k+2)^{2} \phi^{\frac{-6k}{k+2}} &=& (k+2)^2\times \left[\alpha(\rho-\mathcal{H})+\beta(\rho-p)+\Lambda_{0}\right]\label{eq:14}
\end{eqnarray}
From eqns. \eqref{eq:12}-\eqref{eq:14}, here we have expressed the dynamical parameters, the pressure $p$, matter energy density $\rho$ and the magnetic energy density $\mathcal{H}$ in the form of the scale factor as,

\begin{eqnarray}\label{eq:15}
p &=& \frac{-1}{2\alpha(\alpha+2\beta)}\left[ (\alpha+2\beta)S_{1}+\alpha S_{2}-2\beta S_{3}-2\alpha \Lambda_{0}\right]\\
\rho &=& \frac{1}{2\alpha(\alpha+2\beta)}\left[ \alpha S_{2}-(\alpha+2\beta)S_{1}+2(\alpha+\beta)S_{3}-2\alpha \Lambda_{0}\right]\\ \label{eq:16} 
\mathcal{H} &=& \frac{1}{2\alpha}\left(S_{2}-S_{1} \right)\label{eq:17} 
\end{eqnarray}
where, 
\begin{eqnarray}
S_1 &=& \frac{6}{(k+2)}\frac{\ddot{\mathcal{R}}}{\mathcal{R}}+\frac{3(5-2k)}{(k+2)^2}\frac{\dot{\mathcal{R}}^{2}}{\mathcal{R}^{2}}+ \mathcal{R}^{\frac{-6k}{(k+2)}} \nonumber \\
S_2 &=& \frac{3(k+1)}{(k+2)}\frac{\ddot{\mathcal{R}}}{\mathcal{R}}+\frac{3(2k^{2}+1)}{(k+2)^2}\frac{\dot{\mathcal{R}}^{2}}{\mathcal{R}^{2}}- \mathcal{R}^{\frac{-6k}{k+2}}\nonumber \\
S_3 &=& \frac{9(2k+1)}{(k+2)^2}\frac{\dot{\mathcal{R}}^{2}}{\mathcal{R}^{2}}-(k+2)^{2} \mathcal{R}^{\frac{-6k}{k+2}}\nonumber 
\end{eqnarray}

From eqns. \eqref{eq:15}-\eqref{eq:17}, we can derive the EoS parameter $\omega$ and effective cosmological constant $\Lambda$ as 

\begin{eqnarray}\label{eq:18}
\omega &=&-1+\frac{2(\alpha+2\beta)(S_{1}-S_{3})}{(\alpha+2\beta)S_{1}-\alpha S_{2}-2(\alpha+\beta)S_{3}+2\alpha \Lambda_{0}}\\
\Lambda &=& \frac{\beta}{\alpha}\left[S_{3}-S_{1}\right]+\Lambda_0, \label{eq:19}
\end{eqnarray}
We have also expressed some physical parameters that may provide some interesting aspects on the behaviour of the model. These are, the scalar expansion $\Theta=\Sigma H_{i};  i=x,y,z.$; the shear scalar $\sigma^{2}=\frac{1}{2}\left(\Sigma H_{i}^{2}-\frac{1}{3}\Theta^{2}\right)$ and the average anisotropy parameter, $\mathcal{A}=\frac{1}{3}\Sigma\left( \frac{\Delta H_{i}}{H}\right)^{2}$, where $\Delta H_{i}=H_{i}-H.$

\section{Solution with Hybrid Scale Factor}
The dynamical synthesis of the model can be determined through the physical quantities given in eqns. \eqref{eq:15}-\eqref{eq:17}. The formalism developed for the dynamical parameters \eqref{eq:15}-\eqref{eq:17} can help us to analyse the background cosmology for an assumed dynamics of the universe. Therefore, in this work, we have considered the hybrid scale factor (HSF), where the cosmic expansion is governed through a scale factor of the form $\mathcal{R}=e^{at}t^{b},$ where $a$ and $b$ are positive constant and it assumed  $0<[a,b]<1$. In some previous works, the parameter $b$ has been constrained in the range $(0,\frac{1}{3})$ and the other parameter $a$ is left open \cite{Mishra15}. For this HSF, the Hubble parameter, $H=\frac{1}{3}\left(a+\frac{b}{t}\right)$ and the deceleration parameter $q=-1-\frac{\dot{H}}{H}=-1+\frac{b}{(at+b)^{2}}$. It can be seen that at an early phase of evolution, the deceleration parameter $q\simeq -1+\frac{1}{b}$, when $t\rightarrow 0$ and at late phase of cosmic evolution with $t\rightarrow \infty$, when $q\simeq -1$. To maintain the claim of early deceleration and late time acceleration, the deceleration parameter to be positive at early time and negative at late time. FIG. 1 (left panel) represents the deceleration parameter with respect to redshift for the constrained value of $a=0.695$ and $b=0.085$. FIG. 1(right panel) represents the behaviour of Hubble parameter with redshift. As expected for an expanding universe, $H$ decreases from early to late time of the cosmic evolution.

\begin{figure}[!htp]
\centering
\includegraphics[scale=0.50]{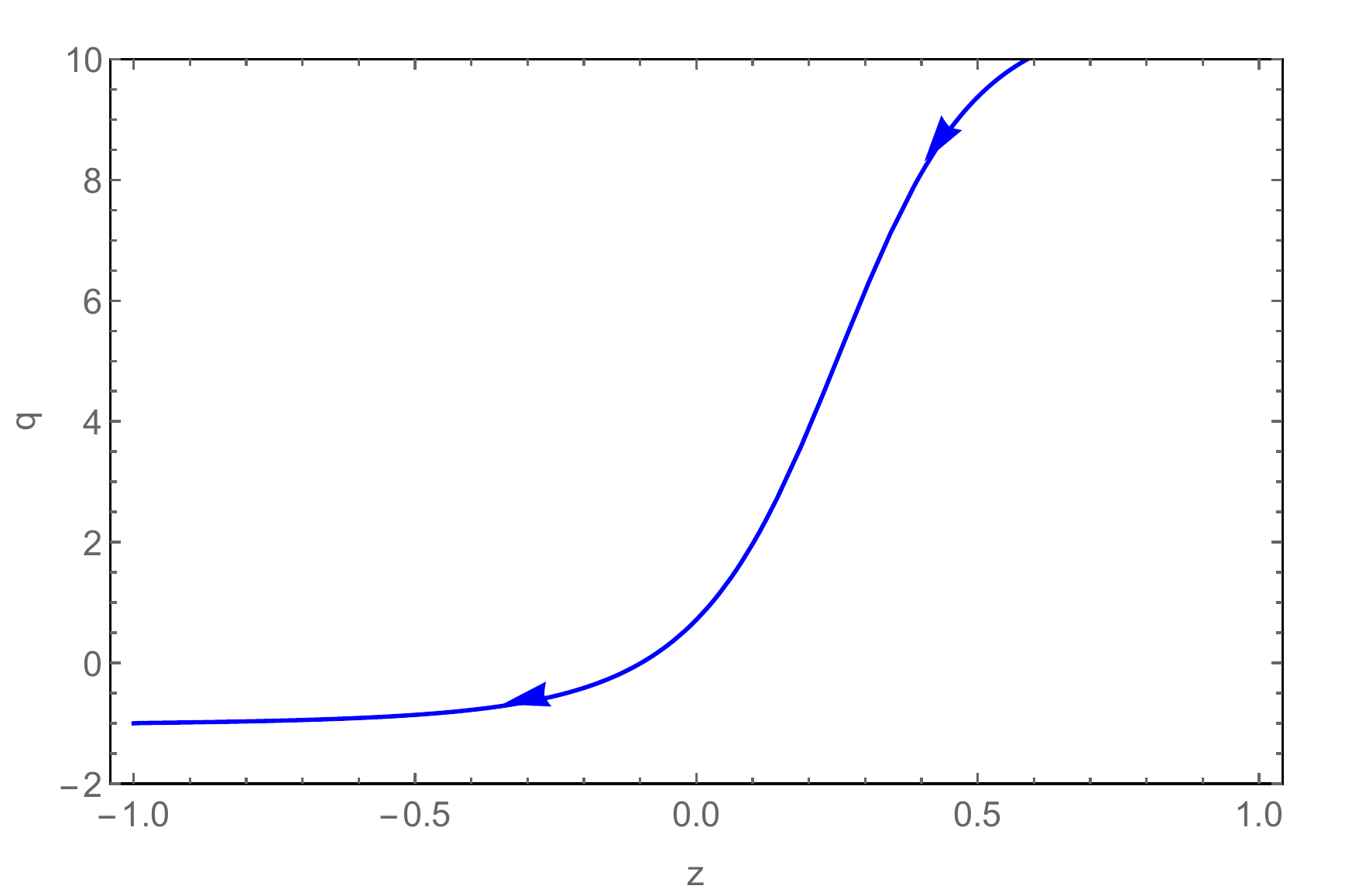}
\includegraphics[scale=0.50]{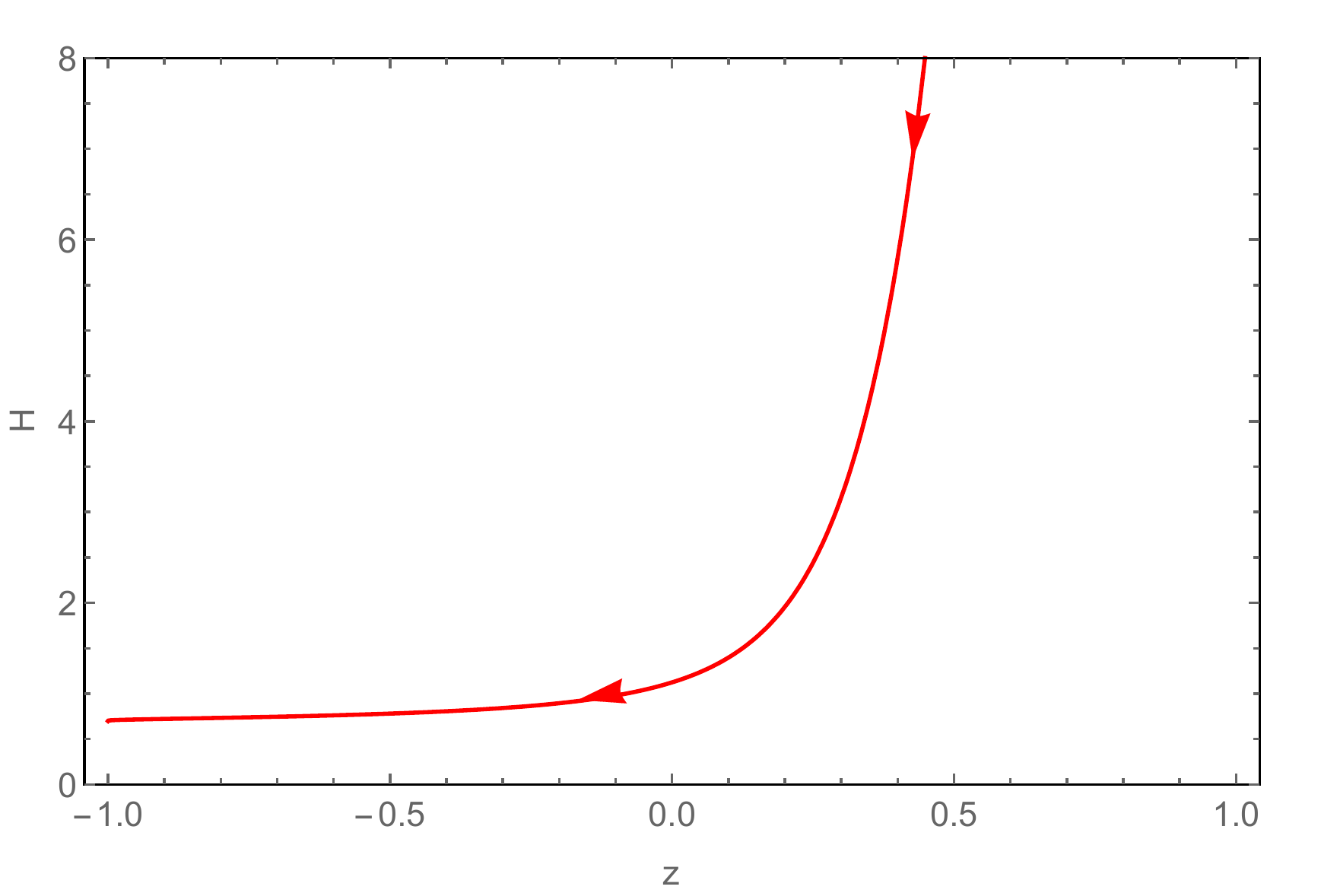}
\caption{Graphical behavior of $q$ (left panel) and $H$ (right panel) versus  redshift function ($z$) for the parametric value $a = 0.695$ and $b=0.085$}
\label{Fig1}
\end{figure}

The scalar expansion and the shear scalar can be obtained for the hybrid scale factor respectively as, $\Theta=a+\frac{b}{t}$ and $\sigma^2=\frac{1}{3}\left(\frac{k-2}{k+2} \right)^{2}\left( a+\frac{b}{t}\right)^{2}$. The graphical representation (FIG. 2) shows that at late time, the expansion remains constant and the shear scalar vanishes. 

\begin{figure}[!htp]
\centering
\includegraphics[scale=0.50]{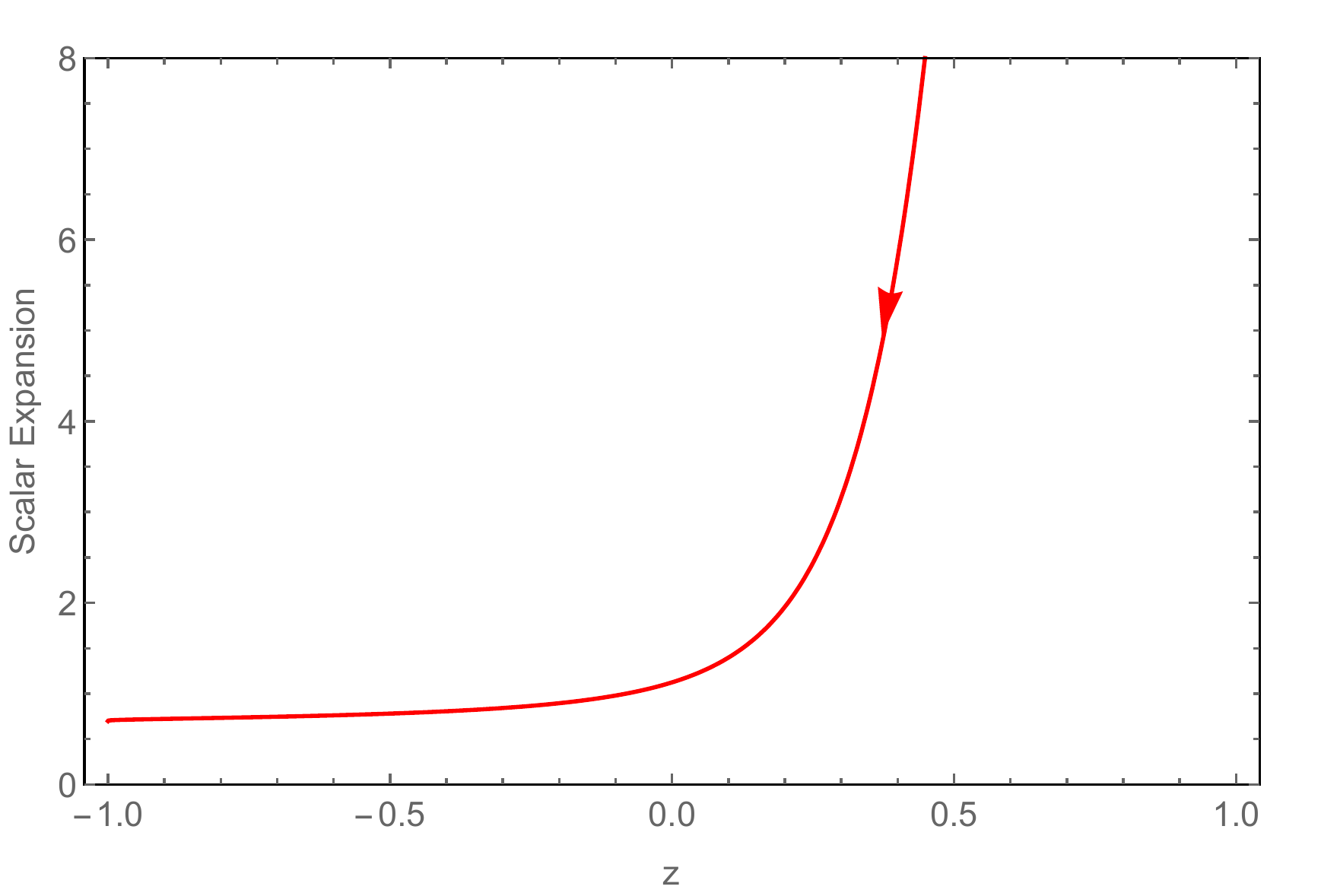}
\includegraphics[scale=0.50]{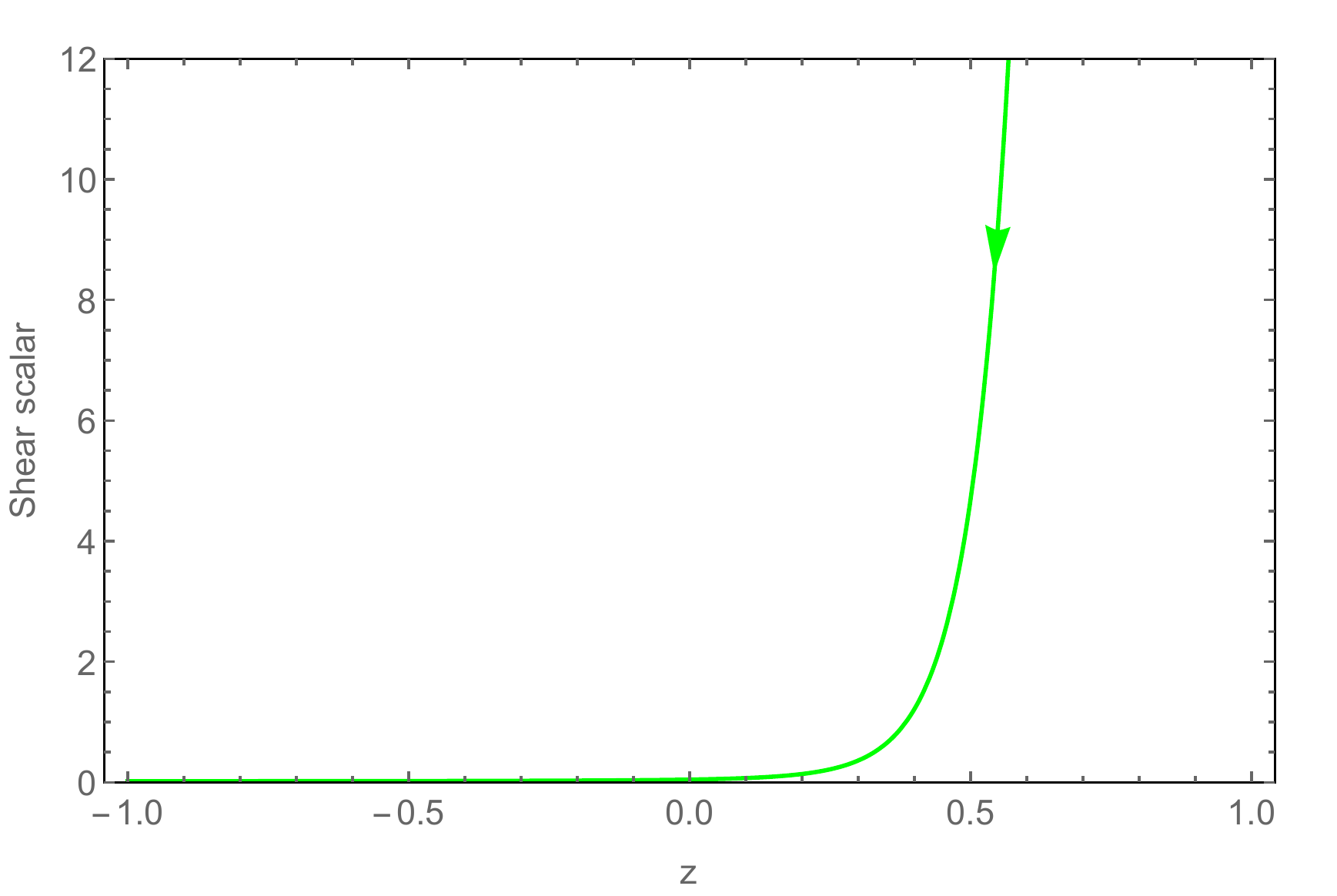}
\caption{Graphical behavior of scalar expansion ($\Theta$) (left panel) and shear scalar ($\sigma^2$) (right panel) versus  redshift function ($z$) for the parametric value $a = 0.695$ and $b=0.085$}
\label{Fig2}
\end{figure}

Introducing this scale factor, the directional scale factors become, $\phi=e^{\frac{akt}{k+2}} t^{\frac{bk}{k+2}}$ and $\varphi=\psi=e^{\frac{at}{k+2}}t^{\frac{b}{k+2}}$. Then eqns. \eqref{eq:15}-\eqref{eq:17} can be reduced to

\begin{eqnarray}
p&=&\frac{-1}{2\alpha(\alpha+2\beta)}\left[ \left( \frac{k_{1}}{(k+2)^{2}}\right)\left( \frac{b^{2}}{t^{2}}+\frac{2ab}{t}+a^{2}\right)-\left( \frac{3\alpha(k+3)+12\beta}{k+2}\right)\frac{b}{t^{2}}+4\beta(e^{at}t^{b})^{\frac{-6k}{k+2}}-2\alpha \Lambda_{0}\right] \label{eq:20}\\
\rho &=& \frac{1}{2\alpha(\alpha+2\beta)}\left[ \left(\frac{k_{2}}{(k+2)^{2}} \right)\left( \frac{b^{2}}{t^{2}}+\frac{2ab}{t}+a^{2}\right)-\left( \frac{3\alpha (1-k)-12\beta}{k+2}\right) \frac{b}{t^{2}}-4(\alpha+\beta)(e^{at}t^{b})^{\frac{-6k}{k+2}}-2\alpha \Lambda_{0}\right] \label{eq:21} \\
\mathcal{H} &=& \frac{1}{2\alpha}\left[\left(\frac{k-1}{k+2}\right)\left(\frac{b^2}{t^2}+\frac{2ab}{t}+a^2-\frac{3b}{t^2}\right)-2 (e^{at}t^{b})^{\frac{-6k}{k+2}} \right] \label{eq:22}
\end{eqnarray}
where, $k_1=3\alpha(3k^2+3k+10)-36\beta(k-1)$, $k_2=9\alpha k(k+5)+36\beta(k-1)$.

\begin{figure}[!htp]
\centering
\includegraphics[scale=0.50]{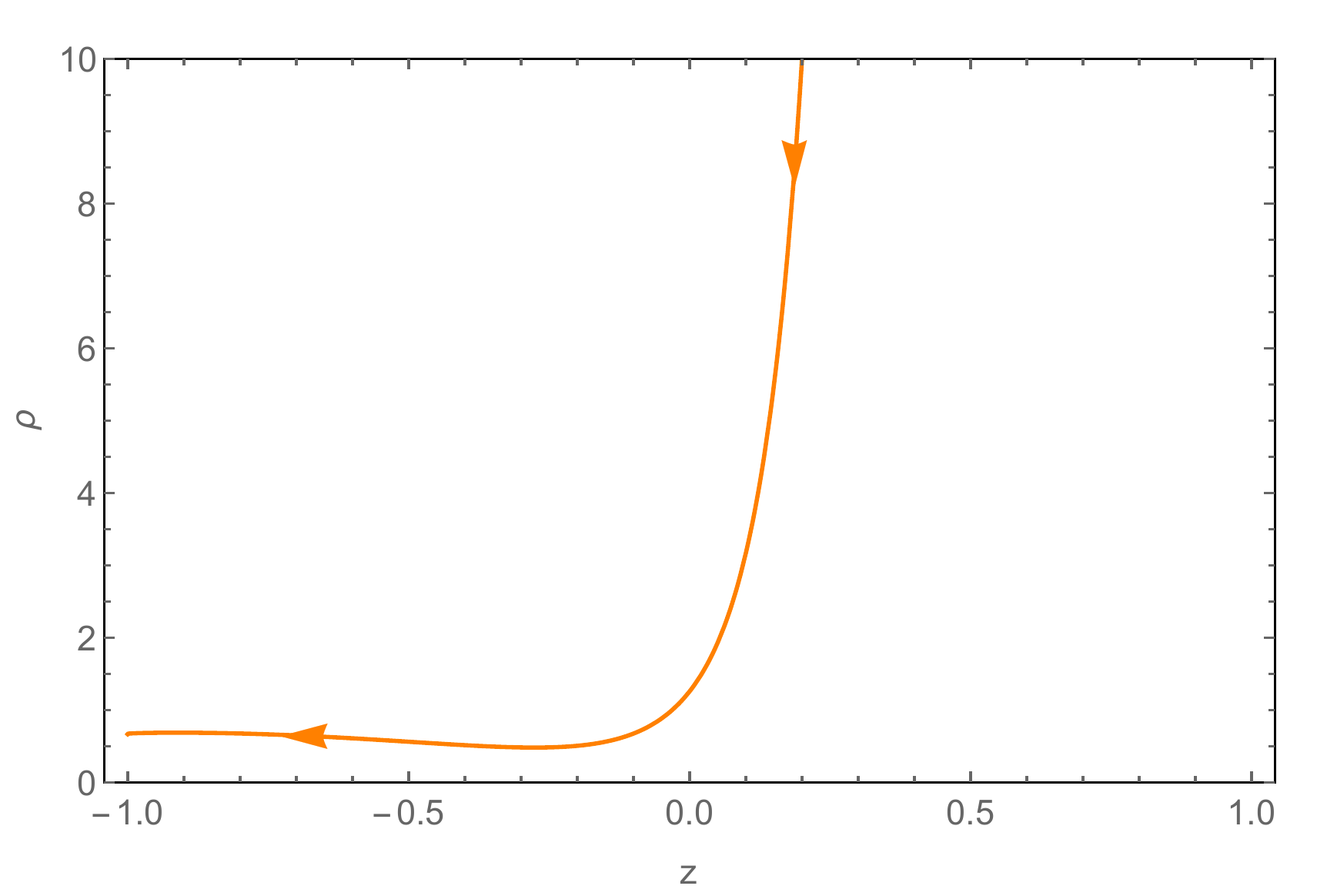}
\includegraphics[scale=0.50]{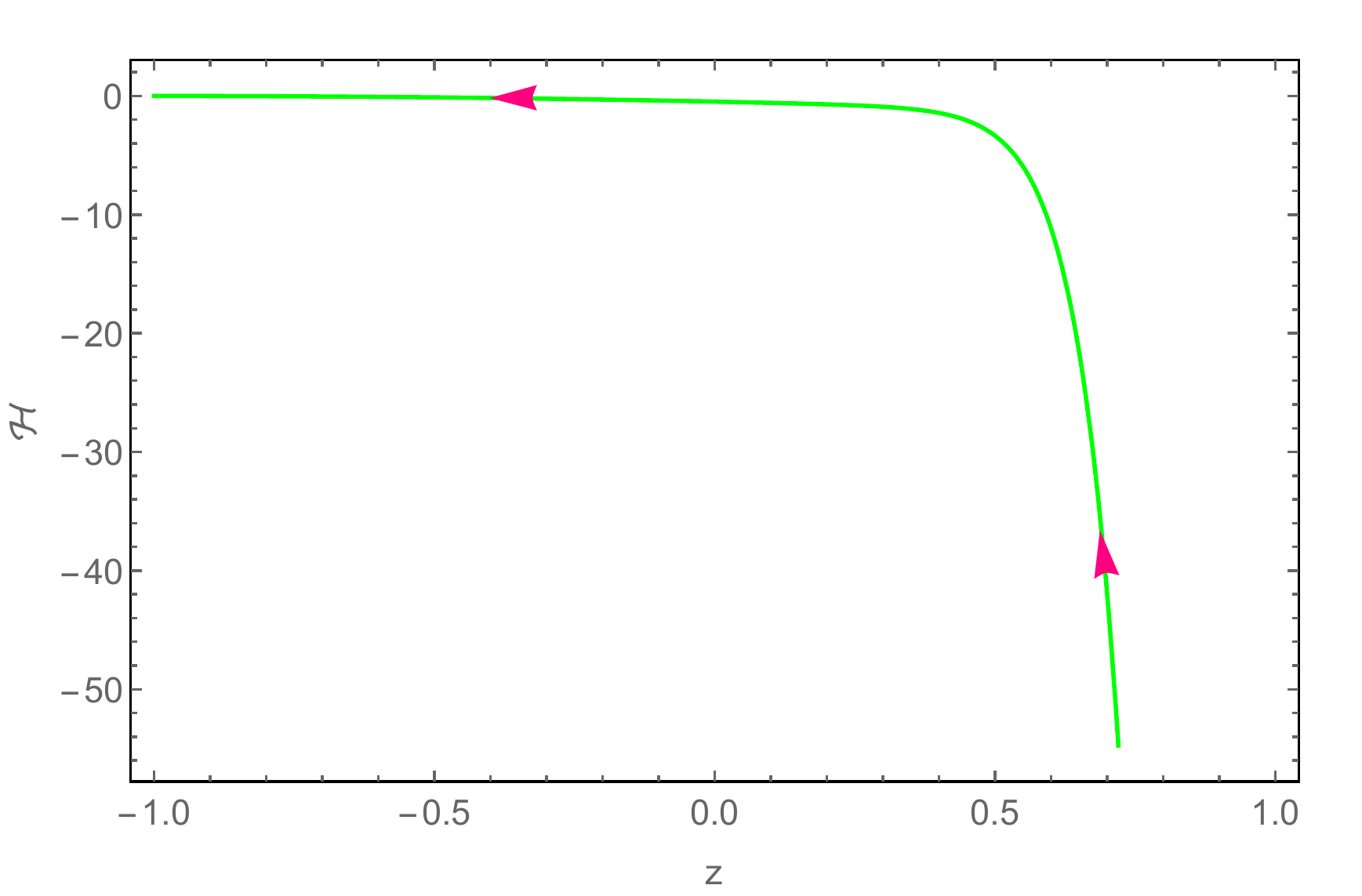}
\caption{Graphical behavior of matter energy density ($\rho$) (left panel) and magnetic energy density ($\mathcal{H}$) (right panel) versus  redshift function ($z$) for the parametric value $a = 0.695$ and $b=0.085$}
\label{Fig3}
\end{figure}

Eqns. \eqref{eq:20}-\eqref{eq:22}, describe the dynamical behavior of the model. The matter energy density and the magnetic energy density are represented in FIG. 3 as in the left panel and right panel respectively. It can be noted that, the physical parameters play a major role for the model and ensures an effective cosmic fluid with positive energy density with negative pressure. One may note from FIG. 3 that, the energy density ($\rho$) being a positive quantity during the cosmic evolution, decreases from a high value at an early time to small value at late time. On the other hand, the magnitude of magnetic field density $\mathcal{H}$ decreases from a large value at an initial epoch to vanishingly small values at late time. This behaviour indicates that, the  magnetic field plays a vital role at an early epoch, possibly providing a strong source for anisotropy.

The EoS parameter $(\omega)$ and the effective cosmological constant $(\Lambda)$ respectively can be obtained as,
\begin{eqnarray}
\omega&=&-1\nonumber \\
&+&\frac{2(\alpha+2\beta)\left((24k-6)\left(\frac{\dot{\mathcal{R}}}{\mathcal{R}}\right)^2-6(k+2)\frac{\ddot{\mathcal{R}}}{R}-2(k+2)^{2}\mathcal{R}^{\frac{-6k}{k+2}} \right) }{\left( 3\alpha(k^{2}+k-2)-12\beta (k+2) \right)\frac{\ddot{\mathcal{R}}}{\mathcal{R}}+\left( 3\alpha(2k^{2}+14k+2)-6\beta(2-8k)\right)\frac{\dot{\mathcal{R}}^{2}}{\mathcal{R}^{2}} -4(\alpha+\beta)(k+2)^{2}\mathcal{R}^{\frac{-6k}{k+2}}-2\alpha(k+2)^{2} \Lambda_{0} }\nonumber \\
\label{eq:23}
\end{eqnarray}
\begin{equation}\label{eq:24}
\Lambda= \frac{\beta}{\alpha}\left[\left( \frac{6(4k-1)}{(k+2)^{2}}\right)\frac{\dot{\mathcal{R}}^{2}}{\mathcal{R}^{2}}-\left( \frac{6}{k+2}\right)\frac{\ddot{\mathcal{R}}}{\mathcal{R}} -2\mathcal{R}^{\frac{-6k}{k+2}} \right]+\Lambda_{0} 
\end{equation}
where $\frac{\dot{\mathcal{R}}}{\mathcal{R}}=a+\frac{b}{t}, \frac{\ddot{\mathcal{R}}}{\mathcal{R}}= \left(a+\frac{b}{t}\right)^2-\frac{b}{t^2}$

\begin{figure}[!htp]
\centering
\includegraphics[scale=0.50]{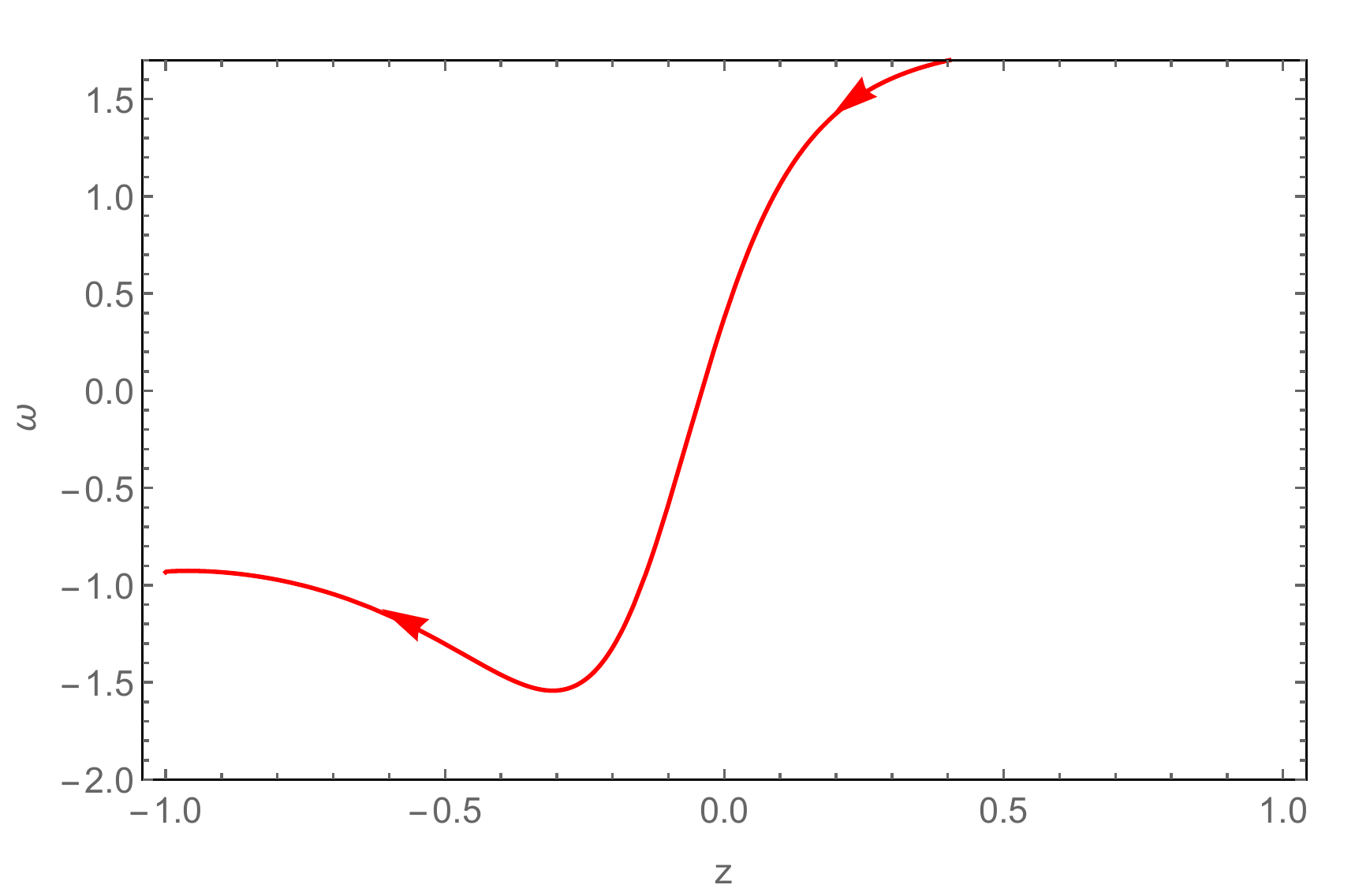}
\includegraphics[scale=0.50]{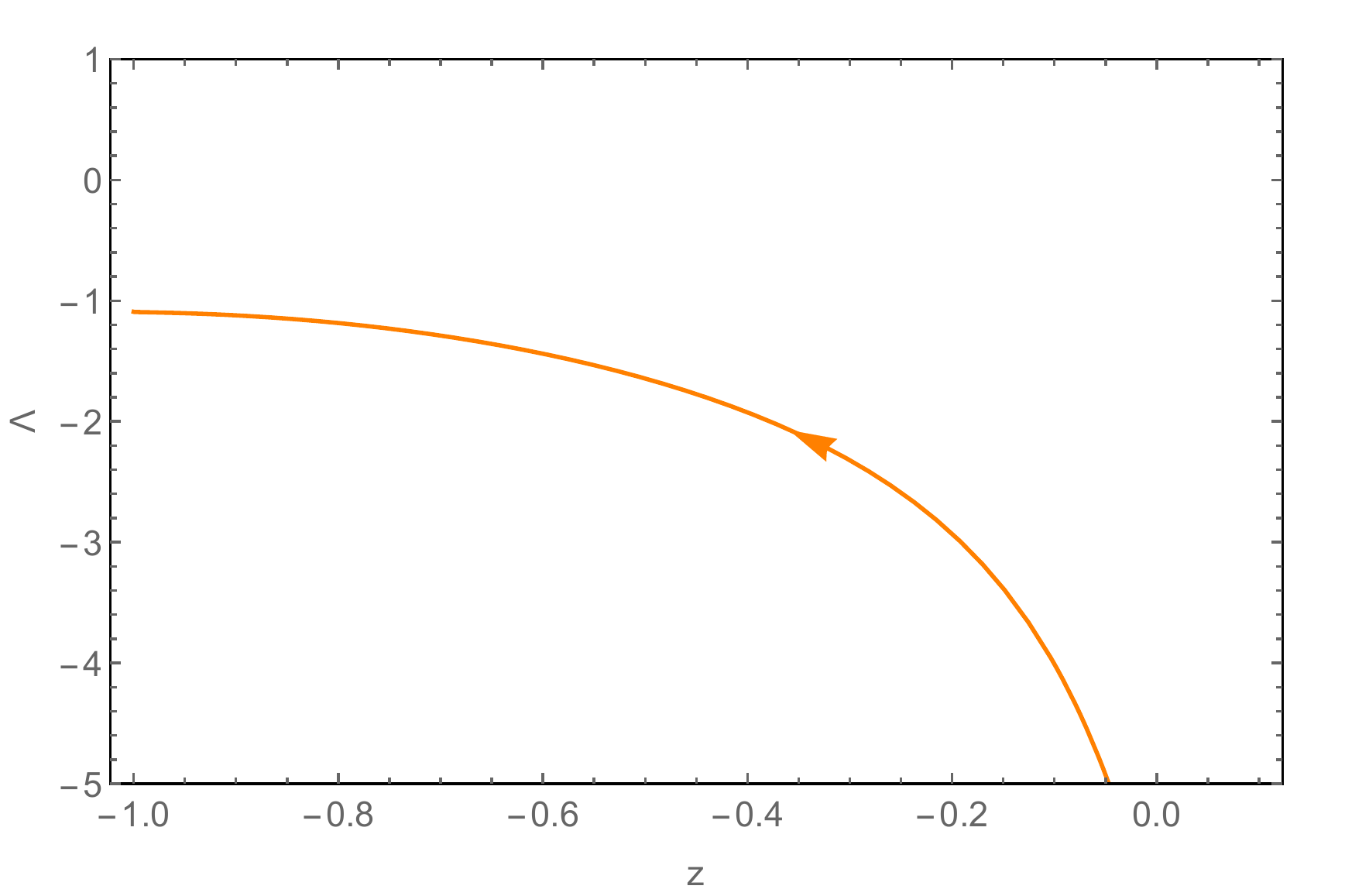}
\caption{Graphical behavior of EoS parameter ($\omega$) (left panel) and effective cosmological constant ($\Lambda$) (right panel) versus  redshift function ($z$) for the parametric value $a = 0.695$ and $b=0.085$}
\label{Fig4}
\end{figure}

We can assess the dynamical behaviour of the model through the evolution of EoS parameter. In FIG. 4 (left panel), the EoS is plotted as a function of the redshift. In plotting the figure,  we have considered $k=1.01$ and $\lambda=2.1, \beta=1.75, \Lambda_{0}=-1.1$. These representative values ensure a positive energy density through out the cosmic evolution in the model. The EoS parameter remains in the negative region and decreases from higher value at the beginning to the lower value at late times. Moreover it passes through the phantom region and at late phase it enters the quintessence region. \\

FIG. 5 (left panel), represents the effect of anisotropy for its representative values  $k=0.90,0.95,1.01$ on the EoS parameter. The coupling constant $\beta=0.75$. It can be seen that, there is no significant changes noticed in the behaviour of the EoS parameter except the fact that, for a lower value of $k$, the EoS parameter goes to a deeper well beyond the phantom divide at some future epoch (around $z=-0.3$). In the right panel of FIG. 5, we have shown the effect of coupling constant $\beta$ on the EoS parameter. Here, we have considered three representative values of $\beta$ namely, $\beta=0.75, 0.95, 1.35$ and kept the anisotropy parameter fixed at $k=1.01$. The evolutionary aspect of the EoS parameter corresponding to a variation of the coupling constant remains almost the same. However, for a higher value of the coupling constant, the model mostly remains in the phantom region. But at the late time it enters into the quintessence region irrespective of the value of $\beta$.\\

\begin{figure}[!htp]
\centering
\includegraphics[scale=0.50]{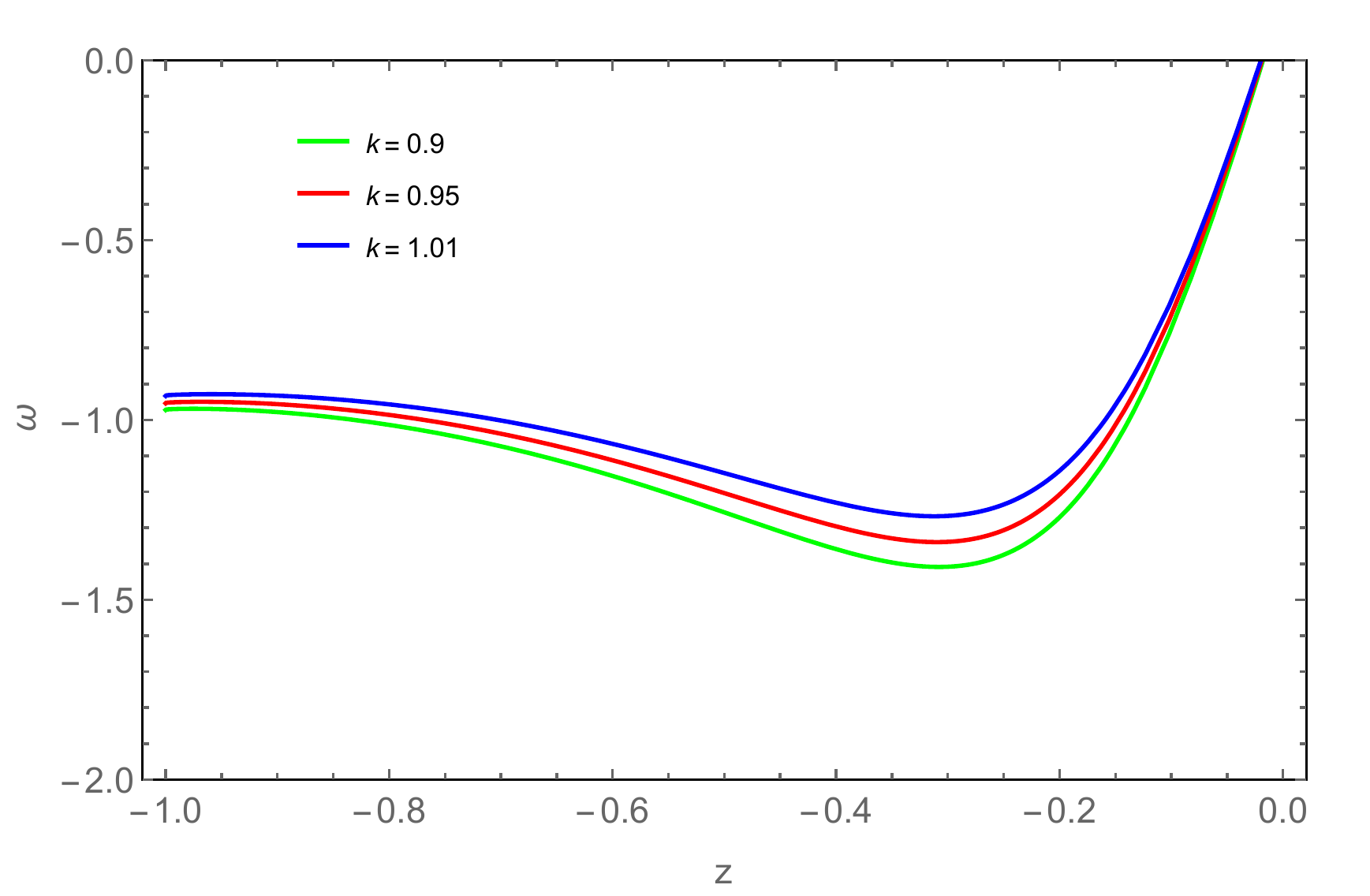}
\includegraphics[scale=0.50]{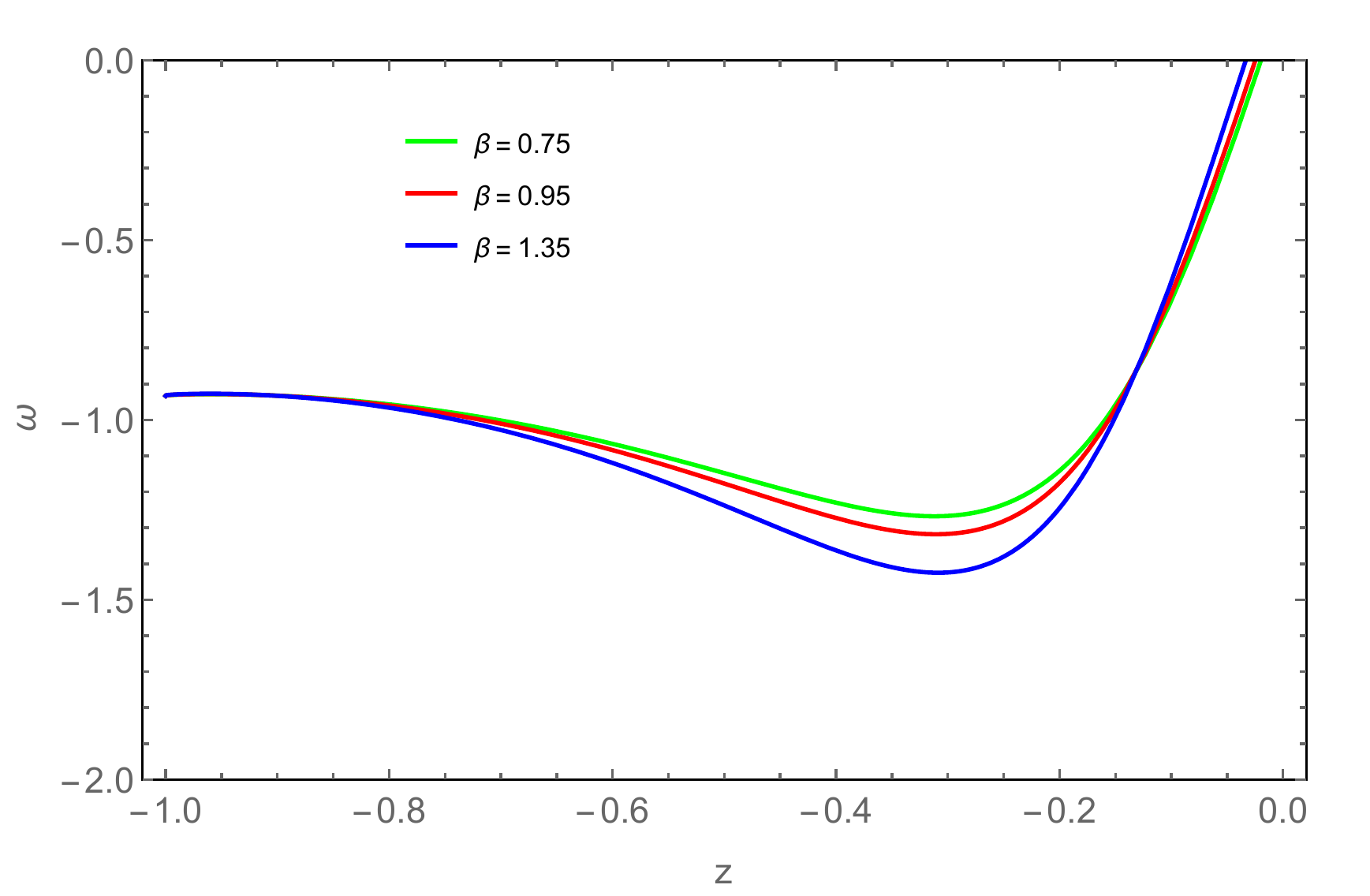}
\caption{Variation of EoS parameter ($\omega$) vs $z$ for, $k=0.90,0.95,1.01$  (left panel) and $z$ for $\beta=0.75,0.95,1.35$ (right panel) }
\label{Fig5}
\end{figure}

From eqn. \eqref{eq:23}, it is evident that the EoS parameter evolves with time that depends on the parameters $\beta$, $k$. We have represented the variation of $\omega$ with respect to the anisotropy parameter $k$ and coupling constant $\beta$ respectively in the left panel and right panel of FIG. 6. For a given value of the coupling constant, the EoS parameter increases with the anisotropy parameter $k$ from a large negative value and converges to zero for higher values of $k$. It can be noted that with smaller value of the coupling constant $\omega$ (Green line), $\omega$ increases faster than the higher value of $\beta$ (Blue line). The transition is recorded at $(-2.967,0.44)$. In the right panel of the FIG.6, we have considered the anisotropy parameter as $k=0.90, 0.95, 1.01$ and plotted the EoS parameter as a function of $\beta$. The plots are linear with negative slopes. It has been observed that the EoS parameter remains  entirely in the negative region and decreases  with the increase in $\beta$. 
  
\begin{figure}[!htp]
\centering
\includegraphics[scale=0.50]{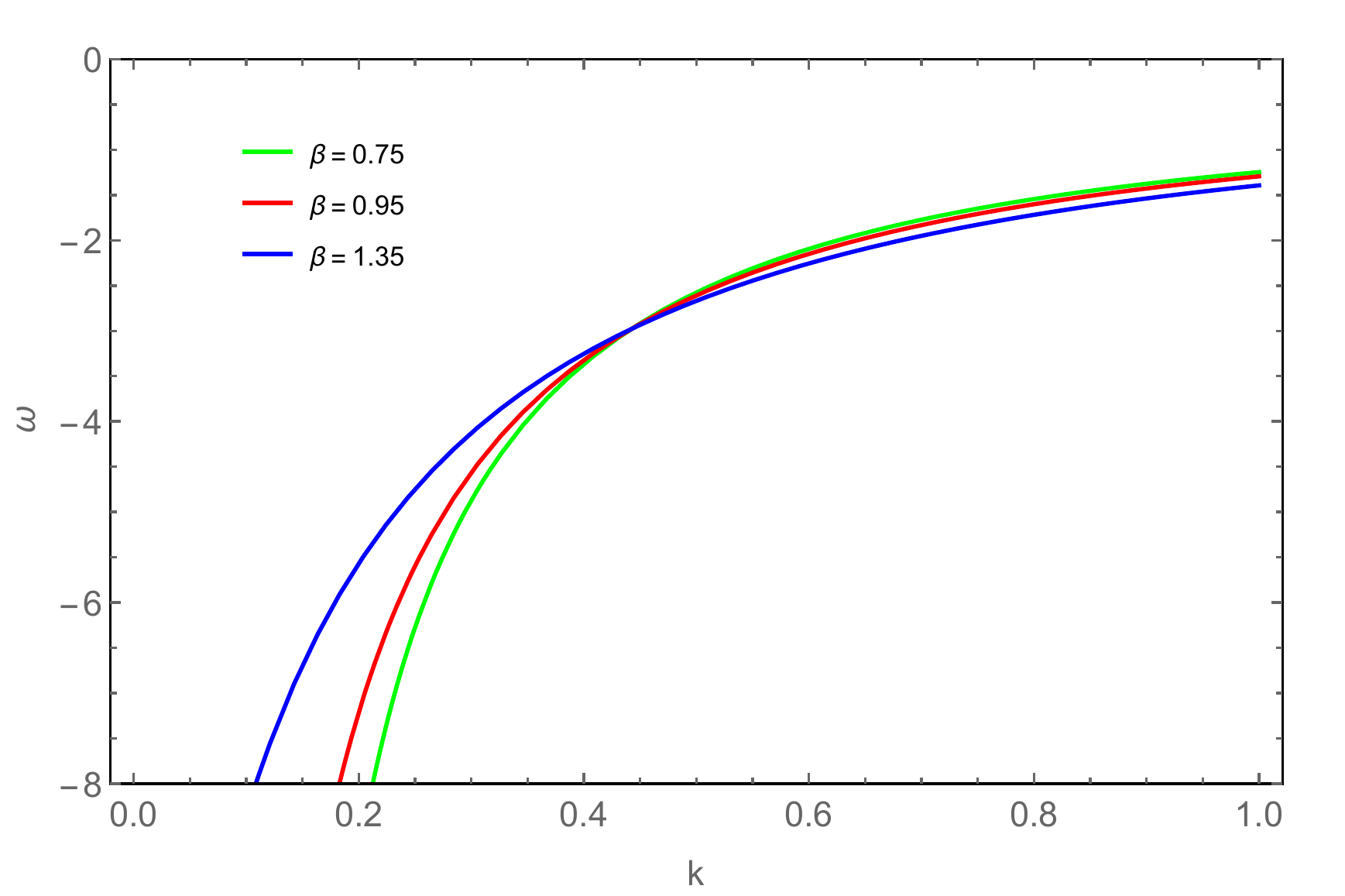}
\includegraphics[scale=0.50]{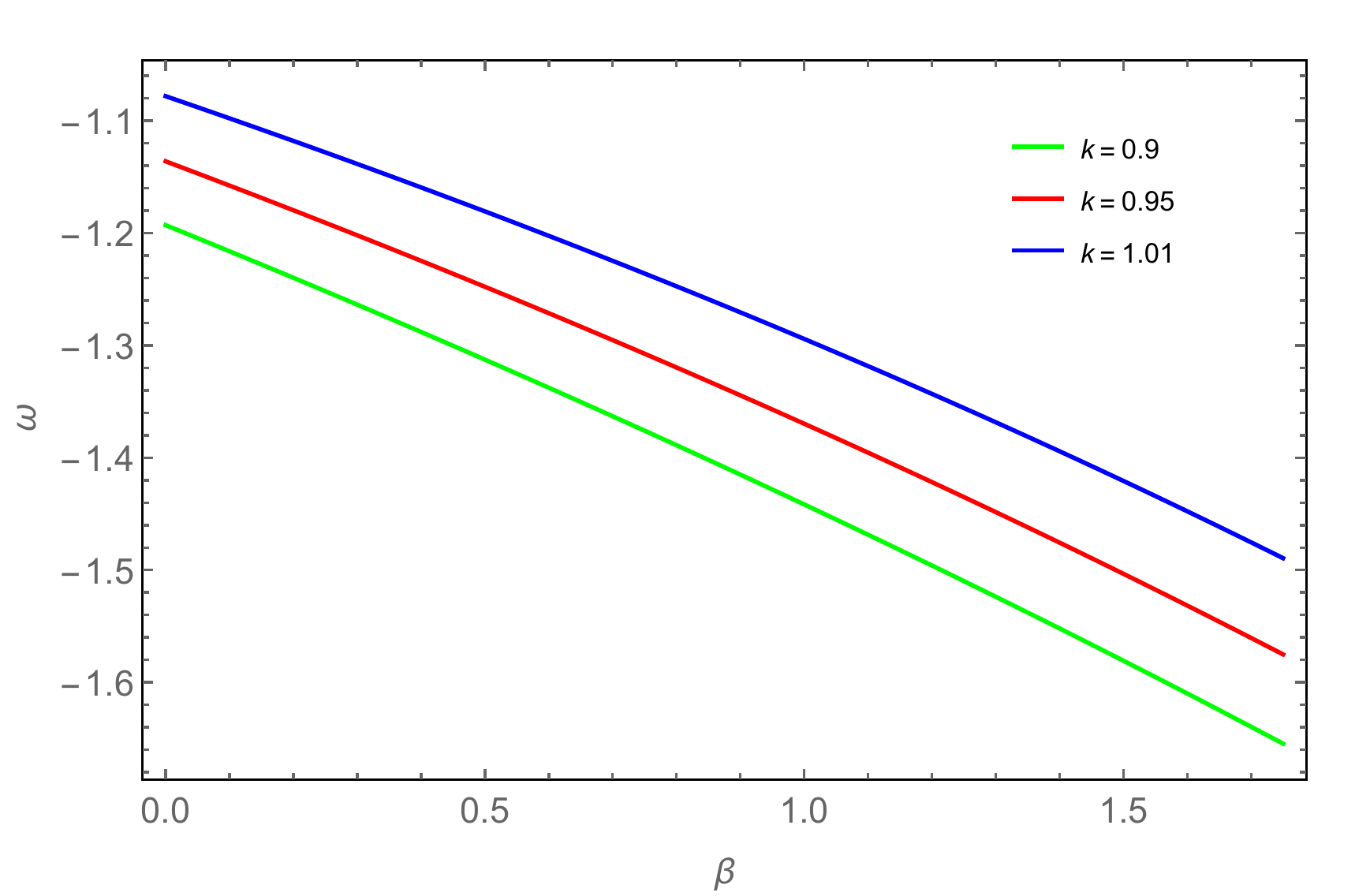}
\caption{Variation of EoS parameter ($\omega$) vs $k$ for $\beta=0.75,0.95,1.35$ (left panel) and vs $\beta$ for $k=0.9,0.95,1.01$ (right panel)}
\label{Fig6}
\end{figure}

\section{Geometrical Analysis and Energy Conditions}

The state finder pair $(r,s)$ is a geometrical analysis of the cosmological model. The geometrical diagnostic is constructed from the scale factor and its time derivative. Through this, we can characterize the properties of the dark energy models. The significant contribution of this pair is that, it can effectively differentiate between different form of dark energy. It is worth to mention here that, from SNAP-type experiment, the mean state finder pair can be determined to very high accuracy \cite{Sahni03}. The state finder pair $(r,s)$ for the hybrid scale factor can be determined as,

\begin{eqnarray}
r&=&\frac{\dddot{\mathcal{R}}}{\mathcal{R}H^{3}}=1-\frac{3b}{(at+b)^{2}}+\frac{2b}{(at+b)^{3}}\\
s&=&\frac{r-1}{3(q-\frac{1}{2})}=\frac{-6b(at+b)+4b}{6b(at+b)-9(at+b)^{3}}.
\end{eqnarray}

At the beginning of the evolution, the state finder pair are (${1 +\frac{2-3b}{b^{2}}, \frac{2}{3b} }$). When $t\rightarrow\infty$, i.e. at the late time $(r,s)$ pair converge to $(1,0)$. So, according to this geometrical diagnostic, the model behaves like $\Lambda_{CDM}$  model at late phase of cosmic evolution. This has been represented graphically in FIG. 7(top panel).\\

Another geometrical diagnostic available in the literature is the $Om(z)$ diagnostic \cite{Sahni08}. The $Om(z)$ parameter is the combination of Hubble parameter and redshift. If the value of $Om(z)$ remains the same for different redshift, then it leads to the cosmological constant model or $\Lambda_{CDM}$ model. The positive and negative slope of $Om(z)$ respectively indicate the phantom($\omega<-1$) and quintessence phase ($\omega>-1$). The $Om(z)$ diagnostic can be obtained for the hybrid scale factor as,
\begin{equation}
Om(z) = \frac{\left( \frac{H(z)}{H_{0}}\right)^{2}-1}{(1+z)^3-1},
\end{equation}
It can be observed from FIG. 7(bottom panel), the model behaves like a cosmological constant model for a substantial time zone in the recent past. Before this period, the model evolves as a phantom field.\\

\begin{figure}[!htp]
\centering
\includegraphics[scale=0.50]{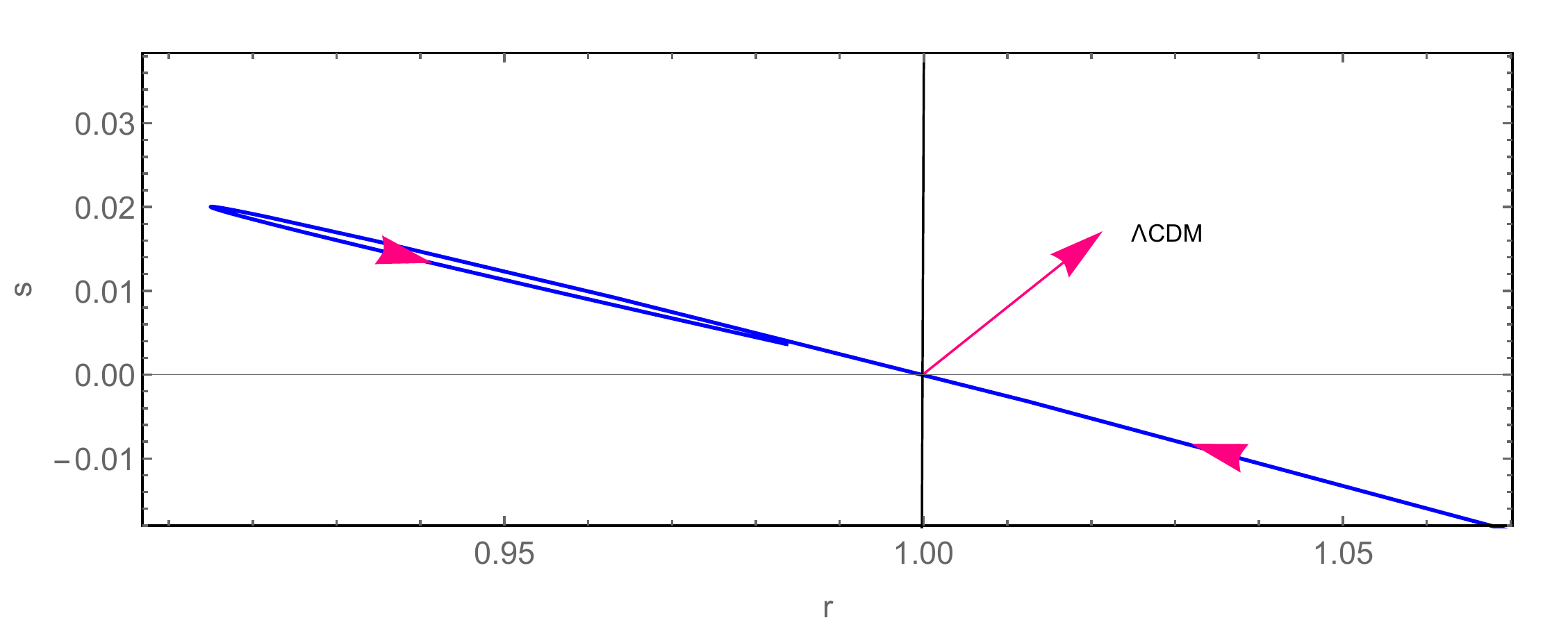}
\includegraphics[scale=0.50]{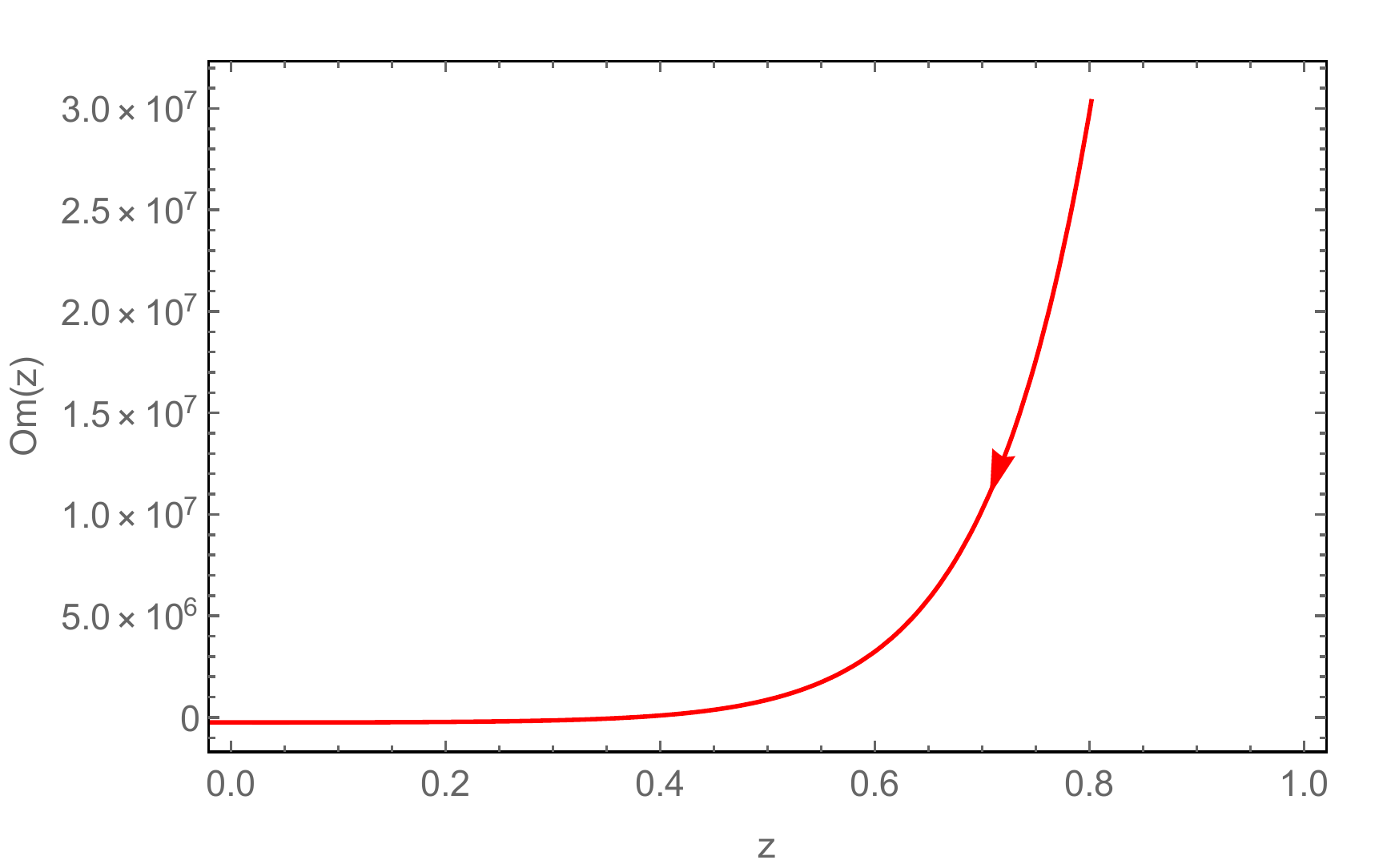}
\caption{Graphical behavior of state finder diagnostic pair(top panel) and $Om(z)$ diagnostic (bottom panel) versus  redshift function ($z$) for the parametric value $a = 0.695$ and $b=0.085$}
\label{Fig7}
\end{figure}

Using eqns. \eqref{eq:20}-\eqref{eq:21}, we will define the null, weak, strong and dominant energy conditions respectively as : (a) $\rho+p \geq 0$,(NEC)  (b) $\rho+p \geq 0,$ $\rho \geq 0$, (WEC) (c) $\rho+3p\geq0$, (SEC) (d) $\rho\pm p\geq0,$ $\rho\geq0$, (DEC). It can be noted that violation of NEC leads to violation of all energy conditions. In $f(R,T)$ gravity with hybrid scale factor, the energy conditions can be calculated as,

\begin{eqnarray} \label{eq:29}
\rho+p&=&\left[\left( \frac{6\alpha(6k-5)}{(k-2)^{2}}\right) \left(\frac{b}{t}+a \right)^{2}+\left( \frac{6\alpha(k+1)+24\beta}{k+2}\right) \frac{b}{t^{2}}-4(\alpha+2\beta)(e^{at}t^{b})^{\frac{-6k}{k+2}}  \right]\times \frac{1}{2\alpha(\alpha+2\beta)} \\
\rho+3p&=&\left[ \left( \frac{k_{3}}{(k+2)^{2}}\right)\left(\frac{b}{t}+a \right)^{2}+\left( \frac{3\alpha(8+4k)+48\beta}{k+2}\right)\frac{b}{t^{2}}-4(\alpha+4\beta) (e^{at}t^{b})^{\frac{-6k}{k+2}}+4\alpha \Lambda_{0} \right]\times\frac{1}{2\alpha(\alpha+2\beta)} \\ \label{eq:31}
\rho-p&=&\left[ \left( \frac{3\alpha(6k^{2}+18k+10)+72\beta
(1-k) }{(k+2)^{2}}\right)\left(\frac{b}{t}+a \right)^{2}-4(\alpha+2\beta)(e^{at}t^{b})^{\frac{-6k}{k+2}}+\left(\frac{12\alpha}{k+2}\right)\frac{b}{t^{2}}-4\alpha \Lambda_{0} \right]\times\frac{1}{2\alpha(\alpha+2\beta)} \nonumber \\ 
& \label{eq:32}
\end{eqnarray}
where $k_{3}=9\alpha(-2 k^2 + 2 k - 10) + 36 \beta (2 k - 2)$.

\begin{figure}[!htp]
\centering
\includegraphics[scale=0.50]{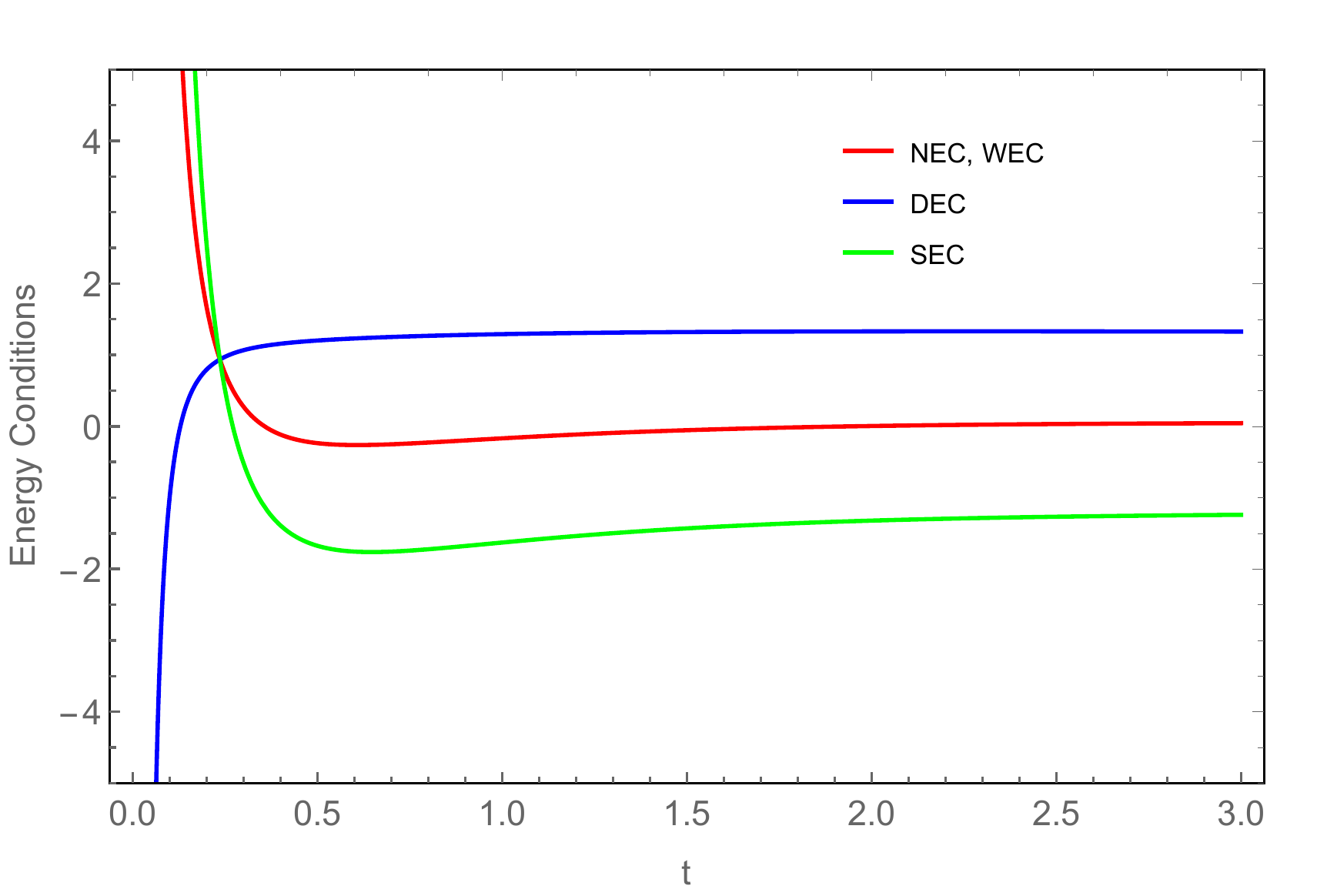}
\caption{Graphical behavior of energy conditions }
\label{Fig8}
\end{figure}

FIG. 8 represents the energy conditions graphically. The energy conditions are observed to change dynamically with the cosmic evolution. The SEC is satisfied in the initial epoch and after $t=0.3$, there is a clear violation of the strong energy condition. On the other hand, the DEC is violated initially and after $t=0.2$, the model satisfied the DEC. However, the NEC and WEC are satisfied in the model. These observations indicate that, the present model in the framework of an extended gravity favours an accelerated expansion of the universe.

\section{Conclusion}
In the present work, we have obtained a magnetized cosmological model in an extended gravity theory which is a special class of the $f(R,T)$ theory. The extended gravity parameters are chosen in such a manner that it can be reduced to the usual GR under suitable conditions. In order to simulate a signature flipping nature of the deceleration parameter and the transitioning aspect of the universe from an early deceleration to an accelerated one, we have considered a hybrid scale factor which behaves as a power law at an early epoch and as an exponential law at late times. The parameters of the hybrid scale factors are chosen so as to reproduce the cosmic transit behaviour at a reasonable epoch. We have considered an anisotropic metric to model the universe. As a source of anisotropy, magnetic field is considered along the symmetry axis. We have developed a formalism to express the field equations and the dynamical parameters of the magnetized model in terms of the Hubble parameter. The dynamical evolution of the EoS parameter is discussed. The EoS parameters evolves from a positive domain in early epoch and cross the phantom divide. Then again, it recovers to evolve into a quintessence phase. The dynamically behaviour of the EoS parameter is marginally affected by the anisotropy parameter considered in the work. However, for a lower value of the anisotropy parameter, it has a deeper well beyond the phantom divide compared to other values. It should be noted here that, in the present work, we have considered the anisotropy parameter that envisages a universe close to isotropic nature. The EoS parameter also changes marginally with the choices of the coupling constant considered in the present work. For a higher value of the coupling constant, it shows a deeper well. However, for a given anisotropy parameter, the behaviour of the EoS parameter is  linear with respect to the coupling constant. From an analysis of the statefinder pair and the $Om(z)$ diagnostic, we found that, the present model behaves substantially as a cosmological constant and overlaps with the $\Lambda_{CDM}$ model  at least at late times.

\section*{Acknowledgement}
BM and SKT acknowledges IUCAA, Pune, India for hospitality and support during an academic visit where a part of this work was accomplished. BM acknowledges DST, New Delhi, India for providing facilities through DST-FIST lab, Department of Mathematics, where a part of this work was done. The authors are thankful to the anonymous referee for the valuable suggestions and comments for the improvement of the paper.

\end{document}